\documentclass[referee]{raa}            % referee version: for submission
\usepackage{graphicx,times}             %for PS/EPS graphics inclusion, new
\usepackage{natbib}
\usepackage{lscape}
\usepackage{graphicx}
\usepackage{enumerate}
\usepackage{amsmath}

\begin{document}

   \title{19 low mass hyper-velocity star candidates from the first data release of LAMOST survey
%\footnotetext{$*$ Supported by the National Natural Science Foundation of China.}
}
%   \subtitle{I. Place Your Subtitle Here}

   \volnopage{Vol.0 (200x) No.0, 000--000}      %%preserved for Editor. DOn't remove!
   \setcounter{page}{1}          %%starting page, preserved for Editor. DOn't remove!

   \author{Yin-bi Li
      \inst{1}
   \and A-Li Luo$^{*}$ \footnotetext{\small $*$ Corresponding Author.}
      \inst{1}
   \and Gang Zhao$^{*}$
      \inst{1}
   \and You-jun Lu
      \inst{1}
   \and Peng Wei
      \inst{1,2}
   \and Bing Du
      \inst{1}
   \and Xiang Li
      \inst{1}
   \and Yong-Heng Zhao
      \inst{1}
   \and Zhan-wen Han
      \inst{3}
   \and Bo Wang
      \inst{3}
   \and Yue Wu
      \inst{1}
   \and Yong Zhang
      \inst{4}
   \and Yong-hui Hou
      \inst{4}
   \and Yue-fei Wang
      \inst{4}
   \and Ming Yang
      \inst{1}
   }

   \institute{Key Laboratory of Optical Astronomy, National Astronomical Observatories,
Chinese Academy of Sciences, Beijing 100012, China;
 {\it lal@nao.cas.cn, gzhao@bao.ac.cn}\\
        \and
             State Key Laboratory of High-end Server \& Storage Technology, Jinan 250510, China\\
        \and
             Key Laboratory for the Structure and Evolution of Celestial Objects, Yunnan Observatories, Chinese Academy of Sciences, Kunming, 650216, China\\
        \and
             Nanjing Institute of Astronomical Optics \& Technology, National Astronomical Observatories, Chinese Academy of Sciences, Nanjing 210042, China\\
   }

   \date{Received~~2009 month day; accepted~~2009~~month day}

\abstract{Hyper-velocity stars are believed to be ejected out from the Galactic center through dynamical interactions between (binary) stars and the central massive black hole(s). In this paper,  we report 19 low mass F/G/K type hyper-velocity star candidates from over one million stars of the first data release of the LAMOST general survey. We determine the unbound probability for each candidate using a Monte-Carlo simulation by assuming a non-Gaussian proper-motion error distribution, Gaussian heliocentric distance and radial velocity error distributions. The simulation results show that all the candidates have unbound possibilities over 50$\%$ as expected, and one of them may even exceed escape velocity with over 90$\%$ probability. In addition, we compare the metallicities of our candidates with the metallicity distribution functions of the Galactic bulge, disk, halo and globular cluster, and conclude that the Galactic bulge or disk is likely the birth place for our candidates.
}

   \authorrunning{Y.-B. Li, A.-Li. Luo \& G. Zhao}             %author_head in even pages
   \titlerunning{Hyper-velocity star candidates from LAMOST}  % title_head in odd pages

   \maketitle
%% The author head (on even pages) and the title head (on odd pages) will be
%% automatically extracted from \author{} and \title{}. Whenever the title is too long,
%% you will be asked to supply a shorter one by inserting either \authorrunning{} or
%% \titlerunning{} before \maketitle. Anyway, you can specify your own heads.
%%
%%
%% Note: In the following text body of your manuscript, please note several differences from
%%       other major journals:
%% (1) \subsection{Please Capitalize the First Letter of Each Notional Word in Subsection Title}
%% (2) Please Capitalize the First Letter of Each Notional Word in all tables' captions

%
%________________________________________________ sections below
%
\section{Introduction}           %% first-level sections will be auto-capitalized
\label{sect:intro}

%% Authors can give a citation as 'Michel et al. 1992'.
%% You may also use \cite, \citep and \citet for citation, and use Table~1 or Figure~1
%% and so forth. Using \ref and \label for cross-references of Tables/Figures
%% is a good way in adjusting/adding/removing text, tables or figures.

The hypervelocity stars (HVSs), discovered in the Galactic halo, are travelling so fast that they can escape from the Galaxy. \citet{1988Natur.331..687H} first predicted their existence, and the existence are the powerful evidence of massive black hole at the center of the Milky Way. A natural explanation is that they may be ejected out from the Galactic center (GC) by interactions of stars with the massive black hole (MBH) or the hypothetical binary MBHs as predicted by \citet{1988Natur.331..687H} and \citet{2003ApJ...599.1129Y}. Such ejection mechanisms can be divided into three categories: tidal breakup of binary stars in the vicinity of a single MBH (\citealt{1988Natur.331..687H, 2003ApJ...599.1129Y, 2006ApJ...653.1194B}), and the binary stars are probably injected into the vicinity of the MBH from the young stellar disk in the GC (e.g., \citealt{2010ApJ...709.1356L, 2010ApJ...722.1744Z}) or from the Galactic bulge (\citealt{2009ApJ...690..795P, 2009ApJ...698.1330P}); single star encounters with a binary MBH (\citealt{2003ApJ...599.1129Y, 2007MNRAS.379L..45S, 2006ApJ...648..976M}); or single star encounters with a cluster of stellar mass black holes around the MBH (\citealt{2008MNRAS.383...86O}).

However, the black hole acceleration mechanism cannot explain a type of HVSs such as US 708 (Hirsch et al. 2005), which is not originated from the center of our Galaxy (e.g., Geier et al. 2015). This type of HVSs is likely to be the ejected donor remnant of a thermonuclear supernova in the white dwarf + helium star scenario (see Wang \& Han 2009; Geier et al. 2015), in which LAMOST are searching for such type of HVSs based on the theoretical results of Wang \& Han (2009).

Alternatively, other ejection models can also accelerate stars to high speed. For example, binary disruption in the dense stellar clusters such as Galactic disk (\citealt{1961BAN....15..265B, 1993ASPC...45..239L, 2012MmSAI..83..272N}), in such case, a supernova explosion of more massive evolved component can accelerate its companion to high speed. Tidal disruptions of dwarf galaxies in the Milky Way can also produce high velocity stars (\citealt{2009ApJ...691L..63A, 2010ApJ...709.1356L, 2009ApJ...707L..22T}), such mechanism can produce high speed star stream or a population of old isolated `escaped' (unbound) or `wondering' (bound) stars.

Seventeen years after Hill's prediction, three HVSs were successively discovered (\citealt{2005ApJ...622L..33B, 2005A&A...444L..61H, 2005ApJ...634L.181E}), they are massive O or B type stars located in the Galactic halo. Until recently, over 20 unbound HVS were identified (\citealt{2009ApJ...690.1639B, 2012ApJ...751...55B, 2014ApJ...785L..23Z}), most of them are massive B type stars, and are located at distant Galactic halo with Galactocentric distances larger than 25 kpc. An interesting exception is a identified B type HVS discovered by \citet{2014ApJ...785L..23Z}, which is the first HVS discovered in the  Large Sky Area Multi-Object Fiber Spectroscopic Telescope (LAMOST) survey, and it is the brightest HVS currently known, and locate at a Galatocentric distance of 13 kpc.

Assuming a Salpeter initial mass function (IMF), the expected solar mass HVSs are about 10 times more abundant than the 3--4 M$_{\sun}$ HVSs (\citealt{2009ApJ...690.1639B}). \citet{2009ApJ...697.1543K} systematically searched for such low mass HVSs in about 290,000 spectra of the Sloan Digital Sky Survey (SDSS) (York et al. 2000; Gunn et al. 1998, 2006), however, they found only 6 metal-poor stars that can be possibly taken as HVS candidates. \citet{2012ApJ...744L..24L} also searched F and G type HVSs from over 370,000 stellar spectra of data release seven of the SDSS, and they presented a low mass metal-poor HVS candidate catalogue. \citet{2014ApJ...780....7P} identified 20 G and K type HVS candidates from approximately 240,000 stars of the Sloan Extension for Galactic Understanding and Exploration (SEGUE) (Yanny et al. 2009) G and K dwarf samples. Besides, \citet{2014ApJ...789L...2Z} reported a catalog of 28 high velocity star candidates from the first data release (DR1) of the LAMOST general survey (Luo et al. 2015), which cover a much broader color range than ever, and 17 of them are F, G, or K type low mass stars.  These current searching results might suggest that the IMF of the parent population of these HVSs is top heavy. A top heavy IMF of the HVS parent population is possibly consistent with the disk origination (\citealt{2010ApJ...708..834B, 2013ApJ...768..153Z, 2010ApJ...723..812K}). In order to distinguish the ejection mechanisms of HVSs and put constraints on the origin of the parent population of HVSs, it is quite necessary to search for the low-mass HVSs.

In this paper, we systematically search and investigate HVSs with stellar spectra of LAMOST DR1, and totally find 19 HVS candidates. The structure of this paper is as follows. In Section 2, we introduce the LAMOST and DR1 data productions in detail. In Section 3, we present a series of spectroscopic, photometric and dynamic criteria to select HVS candidates. In Section 4, we analyze the probability for each HVS candidate that they can escape from our Galaxy. In Section 5, we compare the metallicities of our 19 HVS candidates with the [Fe/H] distributions of Galactic bulge, disk, halo and globular cluster, and conclude that our HVS candidates are likely originated from the Galactic bulge or disk. Finally, the discussion and conclusion are given in Section 6.

\section{The first data release of the LAMOST General Survey}
\label{sect:lamost}

The LAMOST is a 4 meter quasi-meridian reflecting Schmidt telescope, it adopts novel active optics technique, which allows both a large effective aperture of about 4 m and a wide field of view of 5$^{\circ}$. The focal surface of LAMOST has 4000 precisely positioned optical fibers, which are equally connected to 16 spectrographs, thus it can observe 4000 targets simultaneously. Each spectrograph is equipped with a red channel CCD camera and a blue one, which can simultaneously provide red and blue spectra of observed targets respectively (\citealt{2012RAA....12.1197C,1996ApOpt..35.5155W,2004ChJAA...4....1S}).

The primary scientific goal of LAMOST survey is to investigate the large-scale structure of the universe, as well as structure and evolution history of Galaxy, and it consists of two main parts. The first part is the LAMOST Extra-Galactic Survey (LEGAS) of galaxies, and the second part is the LAMOST Experiment for Galactic Understanding and Exploration (LEGUE) Survey of the Milky Way (\citealt{2012RAA....12..723Z,2012RAA....12..735D}). Considering the science goals and the targets available, the LEGAS is consists of galaxy survey and QSO survey, and the LEGUE is divided into three parts, i.e., the Galactic anticenter survey, the disk survey, and the spheroid survey (\citealt{2014IAUS..298..310L,2012RAA....12..805C,2012RAA....12..755C}). The Galactic anticenter survey (LSS-GAC) covers a significant volume of the Galactic thin/thick disks and halo in a continuous sky area of $\sim$ 3,400 square degree, which is centered on Galactic anti-center area with Galactic longitudes 150$^{\circ}$ $\leq$ l $\leq$ 210$^{\circ}$ and latitudes $|b|$ $\leq$ 30$^{\circ}$ (\citealt{2014IAUS..298..310L, yuan2014a}). The disk survey choose eight low and bright plates along the Galactic plane which are nearly uniformed distributed in the region 0$^{\circ}$ $<$ $\alpha$ $<$ 67$^{\circ}$ and 42$^{\circ}$ $<$ $\delta$ $<$ 59$^{\circ}$ (\citealt{2012RAA....12..805C}). The halo survey mainly focuses on areas of SDSS survey, which plan to observe 5.8 million objects with $r <$ 16.8 at $|b|$  $>$ 30$^{\circ}$ (\citealt{2012RAA....12..781Y, 2012RAA....12..735D, 2012RAA....12..792Z}).

After one year pilot survey (\citealt{2012RAA....12.1243L}), LAMOST began its first year general survey from September 2012, and completed its first data release (DR1) to domestic data users and foreign collaborators in June 2013. DR1 totally published 2,204,696 wavelength-calibrated and relative flux-calibrated spectra, which are consists of 1,944,329 stars, 12,082 galaxies, 5,017 quasars, and 243,268 unknown objects. These spectra cover a wavelength range of 3690$-$9100 $\mathrm{\AA}$ with a resolution of R $\sim 1800$. In addition, DR1 also published five spectroscopic parameter catalogs, which are the General catalog, the A, F, G and K type stars catalog, the A type stars catalog, the M dwarfs catalog and the observed plate information catalog respectively. The A, F, G and K type stars catalog provides effective temperatures ($Teff$), surface gravities (log$g$), metallicities ([$Fe/H$]) and heliocentric radial velocities for 1,061,918 stars, these spectroscopic parameters are indispensable for searching and studying hyper-velocity stars (\citealt{Luo2014}).

\section{Identification of the LAMOST DR1 HVS candidates}
\label{sect:identification}
Our HVS candidates are drawn from F, G and K dwarfs of the LAMOST DR1, we use five steps to search for them, and the numbers in below brackets indicate the number of stars left after each step.

\begin{enumerate}
     \item Selecting F, G and K dwarfs from LAMOST DR1[519,027]. The LAMOST DR1 officially  released atmospheric parameters for 1,061,918 A, F, G and K type stars, which are derived by the LAMOST Stellar Parameter Pipeline (LASP) (\citealt{Luo2014}). We fist select F, G and K type stars with 3600 $\leq$ Teff $\leq$ 7500, and plot their Hertzsprung-Russell diagram in Figure~\ref{fig£ºFig1}. From Figure~\ref{fig£ºFig1}, we can see that F, G and K dwarfs can be initially selected with the criteria: 1) log$g >$ 4.0, when 3600 $\leq$ Teff $<$ 6000; 2) log$g >$ 3.75, when 6000 $\leq$ Teff $\leq$ 7500, and we totally obtain 737,023 F, G and K type dwarfs. Then, we upload  equatorial coordinates of our 737,023 dwarfs to the `MyDB' database of SDSS DR10, and obtain photometry information of 521,618 dwarfs from the `PhotoObjall' table, as well as proper motions of 519,027 dwarfs from the `ProperMotions' table. Finally, 519,027 F, G and K dwarfs with both photometries and proper motions are selected.

   \begin{figure}
   \centering
   \includegraphics[width=15cm, angle=0]{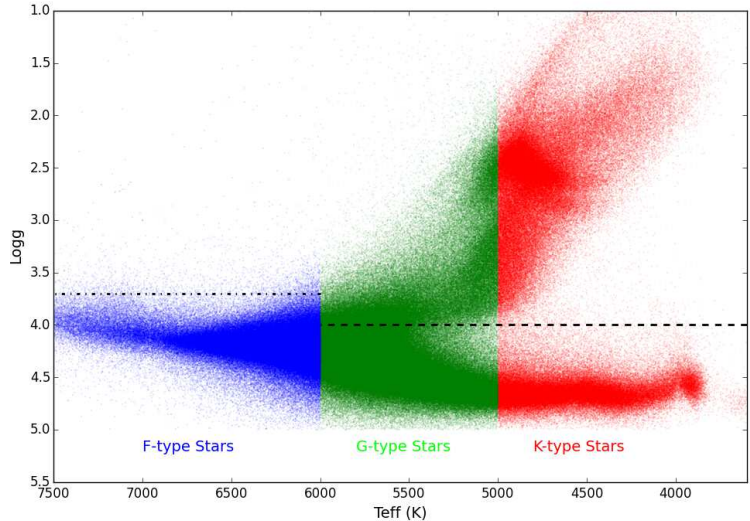}
   \caption{The Hertzsprung-Russell diagram of F, G and K type stars of LAMOST DR1, where blue, green and red points represent F, G, and K type stars respectively. The black dot-dash line shows the location of log$g$ = 3.75 for F type stars, and the black dashed line displays log$g$ = 4.0 for G and K type stars.}
   \label{fig£ºFig1}
   \end{figure}

     \item Further selecting F, G and K dwarfs with the color and magnitude methods[191,405].

         \begin{enumerate}
           \item $g_{0} <$ 20.2, $r_{0} <$ 19.7, $0.2 < (g-r)_{0} <$ 0.48, for F dwarfs \\
                 $r_{0} <$ 19.7, $0.48 < (g-r)_{0} <$ 0.55, for G dwarfs  \\
                 $r_{0} <$ 19.0, $0.55 < (g-r)_{0} <$ 0.75, for K dwarfs
           \item $A(\rm r) <$ 0.5 mag
           \item $|b| \geq$ 10
           \item $psfMagErr\_g/r/i <$ 0.05 mag, $mode =$ 1 and $clean =$ 1
           \item 0.2 mag $< (g - i)_{0} <$ 4.0 mag
         \end{enumerate}

         \citet{2011ApJ...743..187N} investigated the photometric uncertainties of SDSS, and pointed out that they are constant up to r-band apparent magnitude (r$_{0}$) of 19.7, thus we use above criterion (a) to select F, G and K dwarfs. \citet{2012ApJ...761..160S} mentioned that above and below the Galactic plane undergo small amounts of extinction, this small amount of reddening can affect target selection, so we use criteria (b) to retain F, G, and K dwarfs samples with extinction in r-band (A(r)) less than 0.5 mag (A(r) $<$ 0.5 mag). As we known that Schlegel et al. (1998) map has a limited spatial resolution and fails at low Galactic latitudes, extinction from the Schlegel et al. (1998) map may not represent the true value of targets which may lead to inaccurate distance estimates, thus we use criterion (c) to remove dwarfs with $|b| < $10. In addition, in order to make sure SDSS photometry reliable, we use criterion (d) to constrain SDSS psf mag errors in g, r, and i band lower than 0.05 mag, and the value of photometric flag `mode' and `clean' are 1. In the end, we constrain our dwarf samples to the color range of 0.2 $< (g - i)_{0} <$ 4.0 mag, which is the reliable range to estimate absolute magnitude with the photometric parallax relation of \citet{2008ApJ...684..287I}.
     \item Calculate phase space coordinates and escape velocities[191,405]. We first calculate heliocentric distances for 191,405 F, G, and K dwarfs using the distance modulus:
         \begin{equation}
         \label{eq:eq1}
          d(\rm kpc) = \frac{10^{0.2\times(r_{0} - M_{r})}}{100}
         \end{equation}
         where $d$ is the heliocentric distance in unit of kpc, $r_{0}$ is the r-band deredden apparent magnitude directly from the SDSS DR10 `PhotoObjAll' table, and $M_{r}$ is the r-band absolute magnitude.

         The above step constrains our dwarfs in the color range 0.2 $< (g - i)_{0} <$ 4.0 mag, thus we use the photometric parallax relation proposed by \citet{2008ApJ...684..287I} to estimate $M_{r}$:
         \begin{equation}
         \label{eq:eq2}
          M_{r}((g - i)_{0}, [Fe/H]) = M_{r}^{0}((g - i)_{0}) + \Delta M_{r}([Fe/H])
         \end{equation}
         where $M_r^0(g-i)$ is the color-magnitude relation, $\Delta M_r(\text{[Fe/H]})$ is the absolute magnitude correction, and they can be calculated by the equations (A7) and (A2) of \citet{2008ApJ...684..287I}, respectively. The $(g - i)_{0}$ color can be available from the table 'PhotoObjAll' of SDSS DR10, and the metal abundance [$Fe/H$] can be obtained from the  A, F, G and K type stars catalog of LAMOST DR1.

         Then, we use a right-handed cartesian coordinate system centered on the GC to calculate Galactic three dimension (3D) space positions, the X axis pointing from the Sun to the Galactic center (GC) with the Sun at $x$ = -8 kpc, the Y axis pointing in the direction of rotation and the Z axis pointing towards the Northern Galactic Pole. We use a similar coordinate system to calculate Galactic 3D velocity, and assume that the motion of the local standard of rest (LSR) is 220 km s$^{-1}$, and the velocity of the Sun with respect to the LSR is (11.1 km s$^{-1}$, 12.24 km s$^{-1}$, 7.25 km s$^{-1}$) (\citealt{2010MNRAS.403.1829S}).

         To determine which dwarfs are unbound to the Galaxy, we adopt five different Galactic potential models to estimate escape velocities ($V_{\rm esc}$), i.e., a spherically symmetric convergent model
         (\citealt[hereafter Xue08]{2008ApJ...684.1143X}), a spherically symmetric divergent model
         (\citealt[hereafter Kenyon08]{2008ApJ...680..312K}), two axisymmetric divergent models
         (\citealt[hereafter Paczynski90 and Koposov10]{1990ApJ...348..485P,2010ApJ...712..260K}), and a triaxial convergent model (\citealt[hereafter Gnedin05]{2005ApJ...634..344G}). Beside a simple model of Kenyon08, other four potentials are all three-component bulge-disk-halo models. Xue08, Koposov10 and Gnedin05 adopt a spherical bulge (\citealt{1990ApJ...356..359H}), which is different from the Miyamoto-Nagai bulge (\citealt{1975PASJ...27..533M}) of Paczynski90. Paczynski90, Koposov10, and Gnedin05 use the Miyamoto-Nagai disk(\citealt{1975PASJ...27..533M}), while Xue08 adopt an exponential disk. For the halo component, the four three component models are entirely different (\citealt{1996ApJ...462..563N,1990ApJ...348..485P,2006ApJ...651..167F}). For the convergent Galactic potential model, there exists a true escape velocity, and we use equation \ref{eq:eq3} to estimate $V_{\rm esc}$:
         \begin{equation}
         |v_{\rm esc}(r)| = \sqrt{2 \times |\Phi(r)|}  \label{eq:eq3}
         \end{equation}
         However, for the divergent potential, any HVS with finite space velocity could not really escape from the Galaxy. We thus define unbound stars as ones which can reach $r\geq200$ kpc with $v\geq+200$ km s$^{-1}$ as shown in equation \ref{eq:eq4} (\citealt{2008ApJ...680..312K}):
         \begin{equation}
         |v_{\rm esc}(r)| = \sqrt{2 \times (\frac{1}{2} \times 200^{2} + ||\Phi(r)| - |\Phi(200)||)} \label{eq:eq4}
         \end{equation}
     \item Find HVS candidates with clean or reliable proper motions[32]. In step two, we use a series of criteria to make sure that our dwarf samples have reliable SDSS photometric parameters, which affect heliocentric distance estimate accuracy. In addition, the proper motion distribution of SDSS tends to display large proper motion errors,  we thus need to ensure that proper motions of dwarf samples are real rather than the product of large errors.

         \citet{2004AJ....127.3034M, 2008AJ....136..895M} presented an improved proper-motion catalog, which matched each SDSS point source to the USNO-B catalog (\citealt{2003AJ....125..984M}). The SDSS$+$USNO-B catalog is 90$\%$ complete to $g < 19.7$, and has statistical errors of roughly 3-3.5 mas yr$^{-1}$ and approximately 0.1 mas yr$^{-1}$ systematic errors for each component proper-motion. \citet{2004AJ....127.3034M} also defined a series of criteria to make sure that the proper-motions from SDSS catalog are reliable, these criteria were later revised by \citet{2006AJ....131..582K} for their white dwarf samples, and \citet{2014ApJ...780....7P} used the revised criteria when they search for G and K type HVS candidates from the SEGUE. In this paper, we use the criteria to select HVS candidates with `clean' or `reliable' proper-motions, the `clean' proper-motion is defined as follows:
         \begin{itemize}
         \item {\tt match$=$1}, which represents only a USNO-B object is within 1" radius of the SDSS target.
         \item {\tt sigRA $<$ 525} and {\tt sigDEC $<$ 525}, which means that the proper-motion fit must have rms residuals less than 525 mas in both right ascension and declination directions.
         \item {\tt nFit $\geq$ 6}, which shows at least six observations (SDSS+USNO-B) have been used to determine the proper motion.
         \item {\tt dist22 $>$ 7}, which suggests the distance to the nearest neighbor with $g <$ 22 must exceed 7".
         \end{itemize}
         and the `reliable' proper-motion is defined as follows:
         \begin{itemize}
         \item {\tt match$=$1}, {\tt sigRA $<$ 525} and {\tt sigDEC $<$ 525}.
         \item {\tt nFit $=$ 6} and {\tt dist22 $<$ 7}; or, {\tt nFit $=$ 5} and {\tt dist22 $>$ 7}
         \end{itemize}

         \citet{2006AJ....131..582K} pointed out the contamination rate for a target with a `reliable' proper motion is not larger than 1.5$\%$. Finally, we totally find 32 HVS candidates which can escape from the Galaxy in at least one Galactic potential model mentioned in step three, 15 of them have `clean' proper motions, and other 17 candidates have `reliable' proper motions.
     \item Finally, HVS candidates are selected with high quality spectra and reliable atmospheric parameters[19]. After above four steps, 32 HVS candidates are initially selected from over 190,000 F, G and K dwarf samples, we further visually inspect their LAMOST spectra, and find ten proper-motion `clean' candidates and nine proper-motion `reliable' candidates, which have high quality spectra, and their r-band signal-to-noise ratio `SNR$\_$r' are listed in Table~\ref{Tab1}. \citet{2008AJ....136.2022L} presents the SEGUE Stellar Parameter Pipeline (SSPP), which was used to derive the fundamental stellar atmospheric parameters (Teff, logg, and [Fe/H]) for A,F,G and K type stars using multiple methods. We use the version of the SSPP used for the seventh data release of the SDSS to verify atmospheric parameters for 19 HVS candidates, and conclude that the parameters derived by the SSPP are roughly consistent with those obtained by the LASP.
\end{enumerate}

Here, we finally select 19 HVS candidates, all of them have reliable photometry and stellar atmospheric parameters, over 98.5$\%$ probability of robust proper-motions, high quality spectra and no visual blending. Their fundamental parameters such as equatorial coordinates, r-band dereddened apparent magnitudes and atmospheric parameters, are shown in Table~\ref{Tab1}, and their heliocentric distances, Galactic distances, Galactic total velocities, and escape velocities obtained by five Galactic potential models are listed in Table~\ref{Tab2}.

From Table~\ref{Tab1}, we can see that the value of [Fe/H] errors obtained from the LAMOST parameter catalog seem to be large, even much larger than their true value, such large errors will effect the distance error estimates. From Table~\ref{Tab2}, we can see that the Galactic distance errors are really large, and a fraction of them can reach up to or over 10$\%$. Luo et al. (2015) pointed out that external errors released in the LAMOST parameters catalog are larger than the real measurement errors, because they rescaled the external errors using a ratio. They compared the LASP parameters with high resolution spectra results, SDSS DR9 results, and results from Gao et al. (2015) and LSP3(Xiang et al. 2015), the mean external error is 0.125 dex which is much close to the true error. Considering this mean error as [Fe/H] errors of our 19 candidates, the corresponding Galactic distance errors listed in the brackets of the third column in Table~\ref{Tab2} are really much smaller.

Theoretically, Ivezi{\'c} et al.(2008) presented out that a 1.0 dex [Fe/H] error will bring about a 1 mag absolute magnitude (Mr) error at the median thin-disk metallicity ([Fe/H] = -0.2), and a 0.57 mag Mr error at the median halo metallicity ([Fe/H] = -1.50), which represents the mean LAMOST external [Fe/H] error of 0.125 dex will result in a Mr error of 0.125 mag at most, and an approximately 6$\%$ Galactic distance error.

Besides, we check whether our candidates are new findings. \citet{2009ApJ...697.1543K} present 6 F/G type metal-poor HVS candidates from over 290,000 SDSS stars, \citet{2012ApJ...744L..24L} proposed 13 F and G type metal-poor unbound HVS candidates from SDSS DR7, and \citet{2014ApJ...780....7P} found 20 G and K type unbound HVS candidates in SEGUE G and K dwarf samples from the SDSS DR9. Our 19 HVS candidates are not in the three HVS catalogs by checking equatorial coordinates. Besides, \citet{2014ApJ...789L...2Z} presented a catalog of 28 high-velocity star candidates from LAMOST DR1, they use a velocity criterion $|rv| > 200$ km s$^{-1}$ and $|V_{\rm gt}| \geq 300$ km s$^{-1}$  when selecting HVS candidates, where $rv$ is the heliocentric radial velocity, and $V_{\rm gt}$ is the 3D total velocity, and this criterion prevent our candidates except one to be found. Because the  exceptional candidate is not in the $u-g$ and $r - i$ color ranges where \citet{2014ApJ...789L...2Z} estimate accurate photometric metallicity, it is also included in our HVS candidates catalog. Among their 28 HVS samples, there are 12 stars with spectra type earlier than F type, and other 16 candidates are eliminated by our photometry and proper-motion criteria. In summary, we can conclude that our 19 low mass F/G/K type HVS candidates are new findings.

\begin{table}
\centering
\caption{Fundamental parameters of 19 HVS candidates.\label{Tab1}}
\fns
\tabcolsep 3pt
\begin{tabular}{@{}l *{12}c @{}}
  \hline\noalign{\smallskip}
HVS&Designation&\multicolumn{1}{c}{ra$^{a}$}&\multicolumn{1}{c}{dec$^{a}$}&r0$^{b}$&\multicolumn{1}{c}{SNR\_r$^{c}$}&\multicolumn{1}{c}{rv$_{\odot}$$^{d}$}&
\multicolumn{1}{c}{$\mu_{\alpha}cos(\delta)$$^{e}$}&\multicolumn{1}{c}{$\mu_{\delta}$$^{e}$}&\multicolumn{1}{c}{Teff$^{f}$}
&\multicolumn{1}{c}{logg$^{f}$}&\multicolumn{1}{c}{[Fe/H]$^{f}$}&\multicolumn{1}{c}{[Mg/Fe]$^{g}$}\\
 &&\multicolumn{1}{c}{(degree)}&\multicolumn{1}{c}{(degree)}&\multicolumn{1}{c}{(mag)}&&\multicolumn{1}{c}{($\rm km\ s^{-1}$)}&\multicolumn{1}{c}{($\rm mas\ yr^{-1}$)}&\multicolumn{1}{c}{($\rm mas\ yr^{-1}$)}&\multicolumn{1}{c}{(K)}&&&\\
  \hline\noalign{\smallskip}
1&J172524.12+565709.6&261.3505&56.95267&13.95&101&-103$\pm$11&-17.19$\pm$2.62&88.11$\pm$2.62&5251$\pm$102&4.65$\pm$0.43&-0.76$\pm$0.36&0.43\\
2&J170333.23+373102.3&255.8885&37.51733&17.50&29&-60$\pm$10&-22.60$\pm$3.00&18.79$\pm$3.00&5069$\pm$156&4.40$\pm$0.41&0.08$\pm$0.38&-0.05\\
3&J132422.30+312841.6&201.0929&31.47823&15.76&18&-23$\pm$16&-24.35$\pm$2.36&-27.53$\pm$2.36&5828$\pm$330&4.03$\pm$0.66&0.02$\pm$0.62&0.33\\
4&J091255.48+140413.8&138.2312&14.07052&15.26&13&46$\pm$22&9.82$\pm$2.52&-54.49$\pm$2.52&6231$\pm$307&4.67$\pm$0.36&-0.86$\pm$0.82&0.57\\
5&J130548.65+282410.7&196.4527&28.40298&17.43&16&114$\pm$12&-21.49$\pm$2.81&-37.11$\pm$2.81&5931$\pm$447&4.28$\pm$0.93&-1.61$\pm$1.536&1.11\\
6&J133115.50+150438.9&202.8146&15.07748&18.49&62&-29$\pm$6&-20.87$\pm$3.05&4.08$\pm$3.05&4854$\pm$90&4.68$\pm$0.29&0.12$\pm$0.24&-0.03\\	7&J175513.55+511927.4&268.8065&51.3243&13.87&38&-72$\pm$11&8.06$\pm$2.73&48.98$\pm$2.73&5228$\pm$170&4.45$\pm$0.47&-0.34$\pm$0.46&0.32\\	8&J113116.03+571131.1&172.8168&57.19199&16.19&23&-91$\pm$11&-43.52$\pm$2.64&-38.49$\pm$2.64&6083$\pm$259&4.07$\pm$0.50&-1.47$\pm$0.89&0.58\\	9&J121811.06+284659.9&184.5461&28.78333&14.08&54&-18$\pm$8&-48.94$\pm$2.57&-0.32$\pm$2.57&5068$\pm$117&4.64$\pm$0.35&0.04$\pm$0.30&-0.03\\	10&J115209.12+120258.0&178.038&12.04946&15.86&22&206$\pm$15&-32.11$\pm$2.51&19.24$\pm$2.51&5669$\pm$280&4.11$\pm$0.64&-0.01$\pm$0.57&0.10\\	11&J004028.68+393853.0&10.11953&39.64808&16.13&19&-54$\pm$23&-34.08$\pm$2.57&-23.19$\pm$2.57&6119$\pm$290&4.31$\pm$0.50&-0.51$\pm$0.69&0.42\\		12&J171952.43+525035.6&259.9685&52.84325&15.62&24&-67$\pm$13&-32.84$\pm$3.2&83.20$\pm$3.2&5703$\pm$244&4.20$\pm$0.59&-0.11$\pm$0.51&0.27\\	13&J063934.38+280912.8&99.89327&28.15358&17.02&8&15$\pm$30&0.11$\pm$2.52&15.04$\pm$2.52&6064$\pm$431&3.94$\pm$0.59&-0.33$\pm$0.88&-0.18\\	14&J005233.53+413322.6&13.13972&41.5563&14.81&23&-30$\pm$23&-48.83$\pm$2.55&-32.97$\pm$2.55&6187$\pm$282&4.34$\pm$0.48&-0.43$\pm$0.64&0.02\\	15&J012947.93-021343.2&22.44971&-2.228684&14.85&50&13$\pm$13&-24.22$\pm$3.82&30.44$\pm$3.82&5860$\pm$183&4.25$\pm$0.56&-0.36$\pm$0.42&0.26\\	16&J075303.30+272657.0&118.2638&27.44918&17.81&14&75$\pm$32&-18.13$\pm$3.00&4.51$\pm$3.00&6017$\pm$399&4.26$\pm$0.70&-0.39$\pm$0.75&0.36\\	17&J142235.20+455631.3&215.6467&45.94204&14.89&44&-121$\pm$12&-42.77$\pm$2.43&11.14$\pm$2.43&5231$\pm$138&4.68$\pm$0.43&-0.57$\pm$0.44&0.38\\	18&J130744.34-004449.5&196.9348&-0.747102&17.18&9&14$\pm$20&-9.72$\pm$3.11&11.48$\pm$3.11&6035$\pm$408&4.28$\pm$0.50&-0.27$\pm$0.86&0.56\\	19&J103858.44+565558.1&159.7435&56.93283&13.74&74&-15$\pm$14&-88.10$\pm$5.61&97.36$\pm$5.61&5785$\pm$146&4.28$\pm$0.54&-0.74$\pm$0.41&0.36\\
\hline\noalign{\smallskip}
\end{tabular}
\parbox{120mm}{Notes. The candidates HVS1---HVS10 all have `clean' proper-motions, and the candidates HVS11---HVS19 have `reliable' proper-motions. \\
               $^{a}$ Equatorial coordinate from the SDSS `PhotoObjAll' catalog.\\
               $^{b}$ Dereddened r band apparent magnitude from the SDSS `PhotoObjAll' catalog.\\
               $^{c}$ r-band signal to noise ratio from LAMOST parameter catalog.\\
               $^{d}$ Heliocentric radial velocity from LAMOST parameter catalog.\\
               $^{e}$ Proper motion in both right ascension and declination directions from the SDSS `ProperMotions' catalog\\
               $^{f}$ Atmospheric parameters from the LAMOST parameter catalog.\\
               $^{g}$ Note that there are large uncertainties in [Mg/Fe] measurements.\\
               }
\end{table}

\begin{table}
\centering
\caption{Kinematic parameters of 19 HVS candidates. \label{Tab2}}
\fns
\tabcolsep 2pt
\begin{tabular}{@{}l *{8}c @{}}
  \hline\noalign{\smallskip}
HVS&\multicolumn{1}{c}{d$_{\odot}$$^{a}$}&\multicolumn{1}{c}{R$_{\rm G}$$^{b}$}&\multicolumn{1}{c}{V$_{\rm G}$$^{c}$}&\multicolumn{1}{c}{V$_{\rm esc}$$-$Xue$^{d}$} &\multicolumn{1}{c}{V$_{\rm esc}$$-$Pacyznski$^{d}$}&\multicolumn{1}{c}{V$_{\rm esc}$$-$Koposv$^{d}$}&\multicolumn{1}{c}{V$_{\rm esc}$$-$Kenyon$^{d}$}&\multicolumn{1}{c}{V$_{\rm esc}$$-$Gnedin$^{d}$}\\
 &\multicolumn{1}{c}{(kpc)}&\multicolumn{1}{c}{(kpc)}&\multicolumn{1}{c}{($\rm km\ s^{-1}$)}&\multicolumn{1}{c}{($\rm km\ s^{-1}$)}&\multicolumn{1}{c}{($\rm km\ s^{-1}$)}&\multicolumn{1}{c}{($\rm km\ s^{-1}$)}&\multicolumn{1}{c}{($\rm km\ s^{-1}$)}&\multicolumn{1}{c}{($\rm km\ s^{-1}$)}\\
  \hline\noalign{\smallskip}
1&1.5$\pm$0.2&8.0$\pm$0.2(0.1)&644$\pm$96&491&540&573&592&607\\
2&4.6$\pm$0.9&7.5$\pm$0.9(0.3)&626$\pm$98&500&540&572&603&608\\
3&4.6$\pm$1.5&9.0$\pm$1.5(0.5)&572$\pm$195&493&523&555&595&596\\
4&2.9$\pm$0.9&10.1$\pm$0.9(0.3)&563$\pm$178&474&519&551&574&590\\
5&3.7$\pm$1.4&8.7$\pm$1.4(0.4)&540$\pm$218&492&527&559&593&599\\
6&5.9$\pm$0.8&8.7$\pm$0.8(0.5)&524$\pm$101&505&524&555&608&598\\
7&1.9$\pm$0.4&7.9$\pm$0.4(0.2)&503$\pm$83&492&541&575&594&608\\
8&2.5$\pm$0.6&9.4$\pm$0.6(0.3)&501$\pm$119&481&524&557&581&595\\
9&2.5$\pm$0.5&8.7$\pm$0.5(0.3)&490$\pm$98&488&530&562&589&600\\
10&2.3$\pm$0.7&8.5$\pm$0.7(0.3)&489$\pm$92&489&532&565&590&602\\
11&3.0$\pm$0.9&9.8$\pm$0.9(0.2)&671$\pm$115&476&523&556&576&593\\
12&1.4$\pm$0.4&7.9$\pm$0.4(0.1)&621$\pm$138&492&541&575&593&608\\
13&5.5$\pm$2.3&13.5$\pm$2.3(0.7)&613$\pm$157&451&496&527&548&569\\
14&1.8$\pm$0.5&9.0$\pm$0.5(0.2)&603$\pm$91&482&530&564&582&599\\
15&2.0$\pm$0.4&8.9$\pm$0.4(0.2)&591$\pm$60&484&529&562&585&599\\
16&5.1$\pm$1.7&12.7$\pm$1.7(0.4)&583$\pm$104&457&499&530&554&573\\
17&3.1$\pm$0.6&8.5$\pm$0.6(0.3)&561$\pm$93&491&531&563&592&601\\
18&6.1$\pm$2.6&8.4$\pm$2.6(0.8)&527$\pm$140&506&526&557&609&600\\
19&0.6$\pm$0.1&8.3$\pm$0.1(0.04)&508$\pm$42&488&538&572&589&605\\
  \hline\noalign{\smallskip}
\end{tabular}
\parbox{120mm}{$^{a}$ Heliocentric distances obtained by the distance modulus.\\
               $^{b}$ Galactic distances.\\
               $^{c}$ Galactic total velocities.\\
               $^{d}$ Escape velocities obtained by the Xue08, Pacyznski90, Koposov10, Kenyon08 and Gnedin05 Galactic potential model respectively.}
\end{table}

\section{Estimating the reliability of our candidates with Monte-Carlo method}
\label{sect:reliability}
Although we use a series of criteria to ensure our HVS candidates with reliable photometry, atmospheric parameters and proper-motions in section~\ref{sect:identification}, it is still premature to say that the final HVS samples do not contain false-positive detections. With this in mind, we thus consider the probability that our HVS candidates are in fact unbound to the Milky Way. To obtain such an unbound probability for each HVS sample, we built a Monte-Carlo simulation to sample a million realizations of orbit parameters.

\citet{2011AJ....142..116D} present a non-Gaussian probability distribution function (PDF) of proper-motion errors using quasar samples with `clean' proper-motions, which contains a Gaussian core and an extended wing. Applying this error distribution model, we randomly produce a million total proper-motion errors ($pm_{\rm error}$) with the inverse function method. Assuming the proper-motion error is isotropic in the ra-dec plane, we can produce a million angles `$\theta$' using an uniform distribution model, and obtain a million proper-motion errors in the ra and dec direction using $pmra_{\rm error} = pm_{\rm error} \times \cos(\theta)$ and $pmdec_{\rm error} = pm_{\rm error} \times \sin(\theta)$, where $pmra_{\rm error}$ and $pmdec_{\rm error}$ are random proper-motion errors in ra and dec directions respectively. Using the proper-motion measurements from SDSS in ra and dec directions, two component proper-motion errors from SDSS and the random proper-motion errors $pmra_{\rm error}$ and $pmdec_{\rm error}$, we can obtain a million random two components proper-motions.

In addition, we randomly generate a million radial velocities and heliocentric distances assuming a Gaussian error distribution function, and a million random Galactic 6D phase space coordinates and escape velocities at each Galactic distance in the million realization can be further obtained. In such a Monte-Carlo simulation, unbound probability for each HVS candidate can be derived by the fraction $\frac{N_{V_{\rm gt}>V_{\rm esc}}}{1000000}$, where $V_{\rm gt}$ and $V_{\rm esc}$ are Galactic total space velocities and escape velocities respectively, and $N_{V_{\rm gt}>V_{\rm esc}}$ is the number of $V_{\rm gt}$ larger than $V_{\rm esc}$ in a million realizations.

\begin{figure*}
\begin{center}
\includegraphics[width=8cm,angle=0,clip]{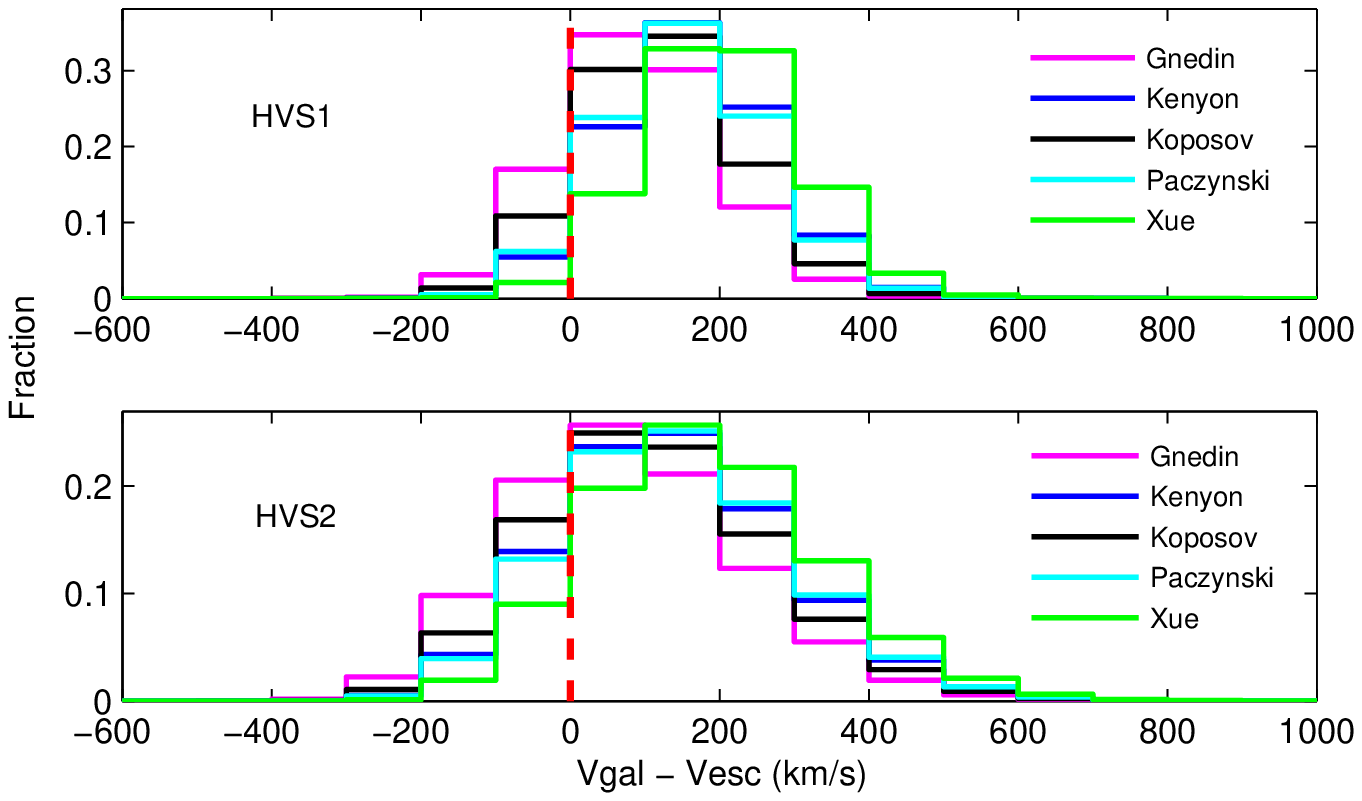}
\includegraphics[width=14cm,angle=0,clip]{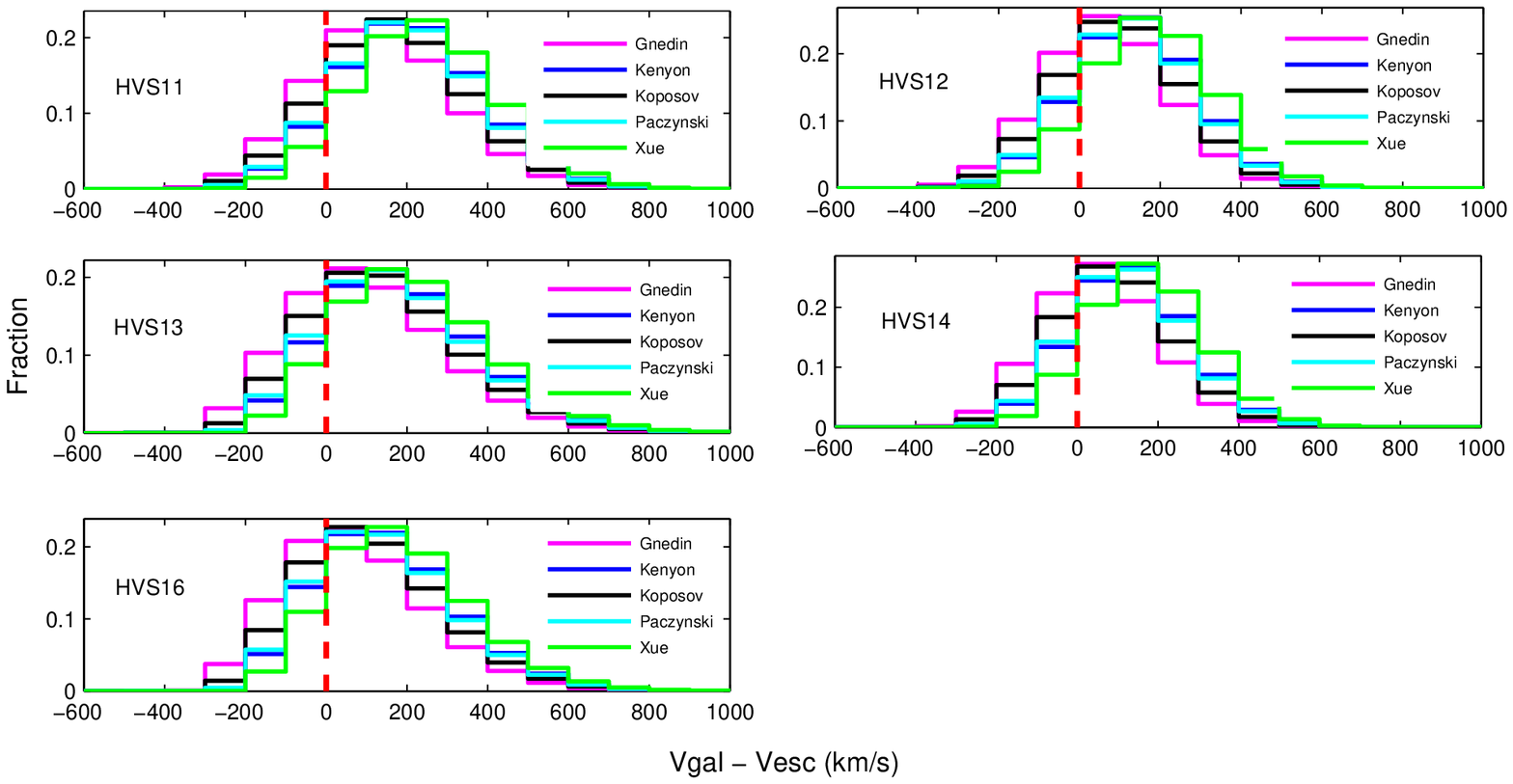}
\caption{The difference distribution of $V_{\rm gal}$ and $V_{\rm esc}$, where $V_{\rm gal}$ and $V_{\rm esc}$ are total Galactic velocities and escape velocities of each sample realization in a million random samples of Monte-Carlo simulation.} \label{fig2}
\end{center}
\end{figure*}

Table~\ref{Tab3} lists unbound possibilities for each HVS candidate, and `---' in this table represents that the candidate is not unbound in this potential model, and we do not calculate unbound possibilities in these cases. From this table, we can see that unbound probability for each HVS candidate exceeds 0.5 as expected, and it is even over 0.9 for HVS1 in the Xue08 potential model. In addition, for each HVS candidate, the value of unbound probability depends on the adopted Galaxy potential model to a certain extent, such as HVS1, the probability varies from 0.64 in the Gnedin05 model to 0.93 in the Xue08 model, and the probability difference between two potential models changes from 0.07 to 0.29. Among our 19 HVS samples, only seven candidates are unbound in all five potential models, Figure~\ref{fig2} shows their distribution of $V_{\rm gal} - V_{\rm esc}$, where $V_{\rm gal}$ and $V_{\rm esc}$ are total Galactic velocities and escape velocities of each realization. From this figure, we can see that the total velocity exceeds escape velocity in most cases.

Actually, the value of unbound probability mainly relies on the difference between total space velocity and escape velocity. When the total velocity is much larger than the escape velocity, the effect of parameter error and Galactic potential model will be extremely small. Conversely, when total velocity is just larger than the escape velocity, parameter error and potential model will greatly affect the unbound probability. Beside, the unbound probability from such a Monte-Carlo simulation can only represents the probability of unbound to the Galaxy when kinematic parameters are in given error ranges.

\begin{table}
\centering
\caption{Probabilities that these candidates are unbound in five Galactic potential models.\label{Tab3}}
\fns
\tabcolsep 2pt
\begin{tabular}{@{}l *{5}c @{}}
  \hline\noalign{\smallskip}
HVS&\multicolumn{1}{c}{P$_{\rm unbound}$-Xue08$^{a}$}&\multicolumn{1}{c}{P$_{\rm unbound}$-Pacyznski90$^{a}$}&\multicolumn{1}{c}{P$_{\rm unbound}$-Koposov10$^{a}$}&\multicolumn{1}{c}{P$_{\rm unbound}$-Kenyon08$^{a}$}&\multicolumn{1}{c}{P$_{\rm unbound}$-Gnedin05$^{a}$}\\
  \hline\noalign{\smallskip}
1&0.93&0.84&0.75&0.86&0.64\\
2&0.80&0.72&0.64&0.70&0.54\\
11&0.87&0.80&0.75&0.82&0.67\\
12&0.80&0.70&0.62&0.71&0.53\\
13&0.81&0.73&0.67&0.75&0.58\\
14&0.80&0.69&0.60&0.71&0.51\\
16&0.77&0.68&0.61&0.70&0.51\\
15&0.89&0.75&0.62&0.77&---$^{b}$\\
3&0.61&0.57&0.52&---&---\\
4&0.64&0.57&0.52&---&---\\
5&0.57&0.52&---&---&---\\
6&0.58&0.52&---&---&---\\
17&0.69&0.58&---&---&---\\
18&0.53&0.50&---&---&---\\
7&0.54&---&---&---&---\\
8&0.54&---&---&---&---\\
9&0.51&---&---&---&---\\
10&0.50&---&---&---&---\\
19&0.63&---&---&---&---\\
  \hline\noalign{\smallskip}
\end{tabular}
\parbox{120mm}{Notes. \\
               $^{a}$ Unbound probability obtained by the Xue08, Pacyznski90, Koposov10, Kenyon08 and Gnedin05 potential model respectively. \\
               $^{b}$ `---' means the candidate is bound in certain potential model, and we do not calculate unbound probability in this case.}
\end{table}

\section{Metallicity distribution and possible origins}
\label{sect:metallcity}
The metallicity distribution investigation of Galactic populations, including the Galactic bulge, the Galactic disk, the Galactic halo, and the globular cluster, indicate each population have a significantly different metal abundance distribution from others. Meanwhile, metal abundance ([Fe/H]) for a star reflects that of the place where it was born. So, metallicity distribution can be used as a tool to explore the origin of our HVS candidates.

\cite{1996AJ....112..171S} measured [Fe/H] for 322 K giants of the Galactic bulge, and present the metallicity distribution function (MDF) for the Galactic bulge. They found that the mean abundance of their K giant samples is $<\rm [Fe/H]> = -0.11\pm0.03$, and over a half of them are in the range of $-0.4<\rm [Fe/H]<0.3$. \citet{2012ApJ...761..160S} derived the MDF of the Galactic disk using 24,270 G and 16,847 K dwarfs from the SDSS SEGUE, different from previous investigations, this work considered observational biases for the first time, and their G and K dwarf samples are the most complete samples in both number and volume. \citet{2013ApJ...763...65A} estimated metal abundance for individual star in SDSS Stripe 82, and presented an unbiased MDF of the Galactic halo. \citet{1997yCat.7202....0H} compiled a catalogue which contains basic parameters of distances, velocities, metallicities, luminosities, colors, and dynamical parameters for 147 globular clusters in the Milky Way, and we obtained the catalog from the website http://vizier.china-vo.org/viz-bin/VizieR?-source=VII/202\&-ref=VIZ55014f467633.

\begin{figure}
   \centering
   \includegraphics[width=\textwidth, angle=0]{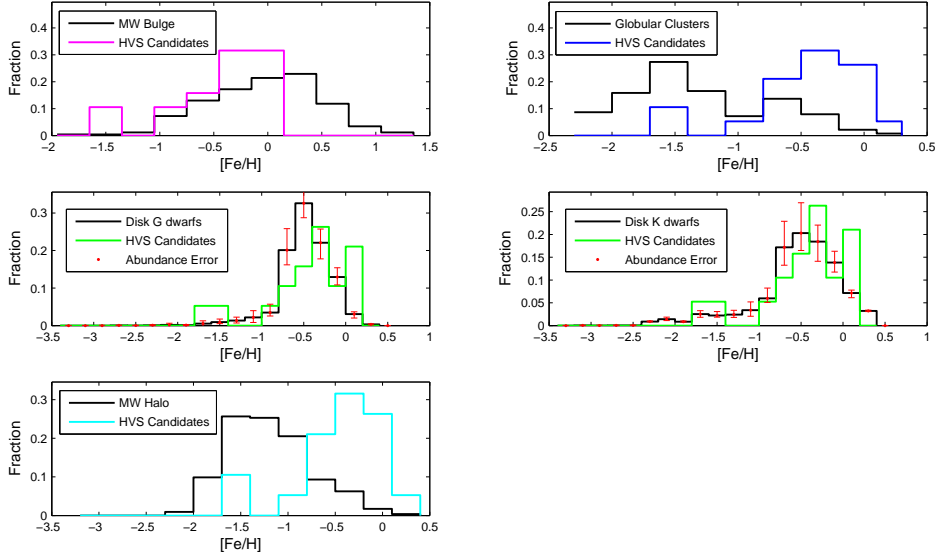}
   \caption{The comparison of metallicity distribution of 19 HVS candidates (black solid line) with the Galactic bulge (magenta), disk (green), halo (cyan) and globular clusters (blue).}
   \label{Fig3}
   \end{figure}

We compare the metallicities of our candidates with the MDFs of the Galactic bulge (\citealt{1996AJ....112..171S}), the Galactic disk (\citealt{2012ApJ...761..160S}), the Galactic halo (\citealt{2013ApJ...763...65A}) and known globular clusters (\citealt{1997yCat.7202....0H}), the MDFs for each Galactic population and our candidates are shown in Figure~\ref{Fig3}. From this figure, we can see that the metallicity distribution of our HVS candidates is well consistent with the G and K dwarf samples in the disk, and is also roughly consistent with the low-metallicity end of the Galactic bulge. Clearly, the MDF of our candidates is completely inconsistent with that of the Galactic halo and globular clusters.

In addition, we estimate Mg abundances [Mg/Fe] using a profile matching method in the region of Mg I b lines around $\lambda5170{\rm \AA}$ (\citealt{2014RAA....14.1423L}), and they are listed in the last column of Table~\ref{Tab1}. The external uncertainty of this profile match method may not be as large as 0.2 dex, and the upper limit of internal uncertainty is 0.3 dex estimated by the Monte-Carlo simulation. From Table~\ref{Tab1}, we can see that the Mg abundances of four candidates, i.e. HVS4, HVS5, HVS8 and HVS18, are larger than 0.5. We visually inspect the spectra of the four candidates, and find that the noise seriously affect spectra quality of the Mg I b region which is extremely important for the [Mg/Fe] estimation. So, we do not regard the four candidates as Mg-enhanced stars. For other candidates, they have high quality spectra in the Mg I b region, and their [Mg/Fe] fall within the range of error. Similarly, the [Mg/Fe] values of our candidates are roughly consistent with the Galactic bulge and disk.

Therefore, our candidates are likely originated from the Galactic bulge and disk, and the Galactic halo and globular cluster may not be their possible origin place. However, the determination of exact birth place need much more reliable parameters, will be released in the future data release of LAMOST, to calculate the intersection regions of our candidates' trajectories and the Galactic disk.

\section{Discussion and Conclusions}
\label{sect:conclusion}
In this paper,  we present 19 F, G or K type hyper-velocity star candidates from over one million stars of the first data release of the LAMOST regular survey. We initially select over half a million F, G and K dwarfs with $Teff$ and log$g$ criteria, and then further pick out over 190,000 final F, G and K dwarf samples with a series of photometric criteria. Then, we obtain 6 D phase space coordinates and escape velocities for each dwarf, and select 17 hyper-velocity star candidates with `clean' proper-motions, and 15 candidates with `reliable' proper-motions. We finally individually inspect spectra of the 32 HVS candidates, and find 19 of them have high quality spectra. Through checking with previous four low mass HVS catalogues in literatures, we conclude that they are all new findings.

Although we use a strict criteria to ensure reliability of the kinematic parameters of our candidates, we still can not confirm that we identify 19 HVS. Therefore, we calculate the unbound probability for each candidate using the Monte-Carlo simulation, assuming a non-Gaussian proper-motion error distribution and Gaussian heliocentric distance and radial velocity error distribution. Such a probability shows each of our candidates can escape from the Milky Way in what extent. We find all the candidates have unbound probabilities over 50$\%$, one of them can even exceed escape velocity with over 90$\%$ probability, and the unbound probability varies in different potential model for each candidate. To investigate the origin of our candidates, we compare metallicities of our candidates with MDFs of the Galactic bulge, disk, halo and globular cluster, and conclude that the Galactic bulge or disk are likely the birth place of our candidates, and the Galactic halo and globular cluster seems not to be their possible origin places.

When we select HVS candidates, there are a large amount of stars with dist22 $< 7$, which implies that they suffer from photometry blending from a near neighborhood. There also exist a large fraction of stars with nFit$<$5, which means that they have few position detections. The Hubble Space Telescope (HST) Fine Guidance Sensor (FGS) and future GAIA can provide more accurate proper-motion measurements and confirm HVS from them, and they can also help us to verify proper motions for our 19 HVS candidates. In addition, high resolution spectroscopic observations are extremely essential to obtain more accurate measurements of stellar atmospheric parameters, which may help us to determine more accurate origin places through calculating trajectories and detailed metallicity distribution analysis, and can also decide whether our HVS candidates are in binaries.

If there exist candidates in binaries, their heliocentric distances and Galactic total velocities will be systematically under-estimated, and their escape velocities will be correspondingly over-estimated, thus these binaries should more likely be able to escape from our Galaxy. Besides, we estimated the effect of binary orbital velocities on the observed heliocentric radial velocities and the Galactic total velocities assuming three types of companion (e.g., a solar mass main sequence companion, a neutron star companion, or a black hole companion) in our previous work (Li, et al. 2012), and we can see that average effect of binary do not exceed 100 km s$^{-1}$, which have little effect on our results.

\begin{acknowledgements}

 We thank an anonymous referee for very useful comments that improved the presentation of the paper. We thank Palladino Lauren E. for valuable discussion of proper motion random sample in unbound probability simulation. This work is supported by the National Science Foundation of China under Grant Nos. 11303036, 11390371/4, 11233004. The Guoshoujing Telescope (the Large Sky Area Multi-Object Fiber Spectroscopic Telescope LAMOST) is a National Major Scientific Project built by the Chinese Academy of Sciences. Funding for the project has been provided by the National Development and Reform Commission. LAMOST is operated and managed by the National Astronomical Observatories, Chinese Academy of Sciences. The web site of LAMOST DR1 is http://dr1.lamost.org.

\end{acknowledgements}

\label{lastpage}

\end{document}